# The benefit of dose-exposure-response modeling in the estimation of dose-response relationship and dose optimization: some theoretical and simulation evidence

Jixian Wang[1], Zhiwei Zhang[2], Ram Tiwari[3]

[1]Bristol Myers Squibb, Switzerland
[2]Gilead Sciences, USA
[3]Bristol Myers Squibb, USA

September 19, 2025

**Abstract**

In randomized dose-finding trials, although drug exposure data form a part of key information for dose selection, the evaluation of the dose-response (DR) relationship often mainly uses DR data. We examine the benefit of dose-exposure-response (DER) modeling by sequentially modeling the dose-exposure (DE) and exposure-response (ER) relationships in parameter estimation and prediction, compared with direct DR modeling without PK data. We consider ER modeling approaches with control function (CF) that adjust for unobserved confounders in the ER relationship using randomization as an instrumental variable (IV). With both analytical derivation and a simulation study, we show that when the DE and ER models are linear, although the DER approach is moderately more efficient than the DR approach, with adjustment using CF, it has no efficiency gain (but also no loss). However, with some common ER models representing sigmoid curves, generally DER approaches with and without CF adjustment are more efficient than the DR approach. For response prediction at a given dose, the efficiency also depends on the dose level. Our simulation quantifies the benefit in multiple scenarios with different models and parameter settings. Our method can be used easily to assess the performance of randomized dose-finding trial designs.

# 1 Introduction

To determine the optimal dose, comparisons of drug responses at different dose levels are fundamentally important [Food and Drug Administration, 2023]. For this purpose, a dose finding trial, in which subjects are randomized to a few fixed dose levels, is the most commonly used one. The mean response at each dose level provides an unbiased estimate due to randomization. However, with small sample sizes in feasible dose-finding trials, there is often high variability in these estimates. Drug exposure data such as trough concentration levels are often measured. Exposure-based analyses are not only useful to achieve precise dose determination, they may also be important to address questions such as dose optimization in special populations, which generally cannot be addressed directly with dose-response (DR) analyzes. Although population pharmacokinetic/pharmacodynamic (PK/PD) modeling is commonly used to predict the response at a given dose, the prediction may be sensitive to misspecification of the dose-exposure (DE) and exposure-response (ER) models, particularly to confounding biases. The use of ER modeling for dose finding has been discussed in general [Hietala et al., 2017; Jones et al., 2019].

To predict the outcome at a given dose level, the sequential PK/PD modeling approach consists of two steps. First, we fit both DE and ER models to the exposure and response data separately. For a given dose, we predict the exposure for individuals using the fitted DE model, then use the predicted exposure in the ER model to predict the response [Wang, 2015]. The mean response can be estimated by averaging the predicted outcomes. This approach will be referred to as dose-exposure-response (DER) approach. In addition to sensitivity to model misspecification, another challenge is the potential confounding bias when fitting the ER model due to confounding factors that affect both the exposure and the response. An approach to diagnose confounders in PK/PD modeling was proposed by Nedelman et al. [2007], and considered in Wadsworth et al. [2018, 2020], in pediatric trials, and reviewed for antibody ER modeling [Kawakatsu et al., 2021].

For phase I dose escalation trials, one early approach using exposure (PK) is to include it as a covariate in the dose-response model [Piantadosi and Liu, 1996] . Recently, several works used DER modeling to guide dose escalation [Ursino et al., 2017; Takeda et al., 2018; Yang and Li, 2023; Yuan et al., 2023] with model setting of different complexity, ranging from simple log-linear to semi-parametric PK models, in combination of ER models such as the Emax model for mean response and a logistic model for toxicity events. Most of these approaches use the classical Bayesian modeling and MCMC to obtain the posterior distribution of, e.g., the probability of event, to guide dose escalation.

Given the extra complexity and sensitivity to confounding biases of the DER approach, one important question is: how much is its benefit, compared with the fitting a dose-response (DR) model to dose and response data directly? Hsu [2009] and Berges and Chen [2013] used simulation to evaluate the benefit of using DER modeling with the Emax ER model. Wang [2015] derived relative efficiency in terms of variance ratio of the two approaches for simple linear ER models. These comparisons assume no unobserved confounding bias. To adjust for this bias, using randomized dose as an instrumental variable (IV) is a useful approach [Nedelman et al., 2007; Wang, 2012]. Intuitively, such an adjustment may reduce the efficiency of the DER approach, hence a similar comparison is useful to evaluate the benefit of the DER approach when confounding adjustment is needed. However, to the best of our knowledge, there has been no comparison for DER modeling with IV adjustment. Our work aims at filling this gap with some theoretical and some simulation results.

In this work, we compare the two approaches in three new settings: 1) when unobserved confounding is adjusted using dose as an IV; 2) when a nonlinear ER model is used; and 3) when some distributional assumptions are dropped. For 2), we focus on ER models with a sigmoid curve. Typical examples are the most commonly used Emax models in pharmacometrics and the logistic model in statistics. In fact, the latter captures the essential non-linearity of the former, since the Emax model can be parameterized as a logistic model. Although the former also allows a flexible range of the response, the essential non-linearity is included in the logistic model. However, neither of the two models is collapsible, that is, a combination of a logistic ER model and a simple linear DE (or PK) model generally does not lead to a logistic DR model. Consequently, a comparison between the DER and DR approaches in terms of their model parameters is not possible. Hence, we use the probit model as a surrogate, since it is known to be very similar to the logistic model, although they appear to be rather different functions. The advantage of the probit model is that when the random terms are normally distributed, a probit DER model leads to a probit DR model. Nevertheless, we also propose an approach without the normality assumption, at the cost of some efficiency loss.

The next section introduces DE and ER models and provides a review of CF approach with discrete IV. The relationship between the (marginal) DR relationship and the DER modeling approach, in particular for probit models, is described in Section 3. Section 4 focuses on using randomized dose as an IV, and CF as a regressor for adjustment. A theoretical comparison between the DER, with and without the CF adjustment, and DR modeling approaches is presented in Section 5. Section 6 presents results of a simulation study for comparison of the two approaches based on probit models. The last section gives a summary and discussion of our findings and some topics that have not been covered for investigation in the future.

## 2 Dose-exposure-response modeling for phase II dose finding trials

We consider the use of DER modeling for phase II dose finding trial design and will focus on the situation of no repeated measurements. Let $D_i, C_i, Y_i$ be the dose, PK exposure (which is often in log scale) and response of subject $i$. Following the potential outcome framework [Rubin, 2005], let $Y_i(d)$ be the potential (counterfactual) response when subject $i$ had dose $D_i = d$, and $Y_i(c)$ be the potential response when subject $i$ had exposure $C_i = c$. We are interested in the DR relationship represented by the mean response as a function of dose $d$: $\mu(d) = E(Y_i(d))$. In general, we can write the DER relationship in terms of the following models

$$\begin{aligned} Y_i &= g(C_i, \epsilon_i) \\ C_i &= h(D_i, \eta_i), \end{aligned} \qquad (1)$$

where $g(.,.)$ and $h(.,.)$ are unknown DE and ER models, $\epsilon_i$ and $\eta_i$ are random variables following some distributions, which at this stage are unspecified. Here, $g(.,.)$ and $h(.,.)$ may be parameterized as $g(C, \epsilon, \beta)$ and $h(D, \eta, \gamma)$, with known functions, but unknown parameters $\beta$ and $\gamma$. A simple parametric form of (1) is given by linear models

$$\begin{aligned} Y_i &= \beta_0 + \beta_c C_i + \epsilon_i \\ C_i &= \gamma_0 + \gamma_d D_i + \eta_i, \end{aligned} \qquad (2)$$

where $\beta_0, \beta_c$ and $\gamma_0, \gamma_d$ are parameters.

The linear DE model can represent some simple popPK models, e.g., the average concentration in steady state with repeated oral dosing interval $\tau$ and the drug PK follows the two-compartmental model with first-order absorption

$$C_{ss} = \frac{FD_{ss}}{Cl\tau} R \qquad (3)$$

where $D_{ss}$ is the steady-state dose, $F$ bioavailability, $Cl = K_e V_d$ with the coefficients of elimination $K_e$ and volume of distribution $V_d$, and $R$ is a non-negative random term. All these parameters, and $R$, are subject or patient specific. A log-transformation makes the above formula a linear model

$$\log(C_{ssi}) = \log(D_{ssi}) + \log(F_i/Cl_i) - \log(\tau) + \log(R_i). \qquad (4)$$

But for simplicity of notation, we denote $C_i = \log(C_{ssi})$ and $D_i = \log(D_{ssi})$ as the exposure and dose on log scale. Also, we let $\eta_i = \log(F_i/Cl_i) + \log(R_i)$ be a random variable on log-scale. Often $\eta_i$ and $\epsilon_i$ are correlated as they may include factors that affect both the exposure and the outcome. For example, elderly patients may have reduced drug clearance, hence a higher $C_i$, and also a higher risk of certain safety events, which may not be related to the drug.

We further assume that $D_i$ are randomized to $K$ levels $d_1, ..., d_K$, and that random errors in the DE and ER models are often correlated. Also, we assume that the marginal DR model can be written as

$$Y_i = g(D_i, \epsilon_i) \qquad (5)$$

where we recycle the notation for the ER model when there is no ambiguity. It can also be parameterized as $g(D, \epsilon, \alpha)$ with known function $g(.,.)$ and unknown parameter $\alpha$.

# 3 Marginal dose-response relationship by dose-exposure-response models

When $D_i$ is randomized, $\mu(d)$ can be estimated using a DR model without exposure data. However, here we consider how to combine the DE and ER models to form a marginal DR model. Although this seems simple, some technical issues are important for the comparison between the DR and DER approaches. An assumption we make here is that the observed $C_i$ completely mediates the drug effect on the response and that the ER model describes the relationship between the observed $C_i$ and the response $Y_i$. If $C_i$ is observed with an error, depending on the nature of the error, the estimation of $\beta_c$ may need an adjustment [Carroll et al., 2006].

## 3.1 Linear models

With the linear DE and ER models (2), the marginal DR model can be derived as:

$$\begin{aligned} Y_i &= \beta_0 + \beta_c C_i + \epsilon_i \\ &= \beta_0 + \beta_c \gamma_0 + \beta_c \gamma_d D_i + \beta_c \eta_i + \epsilon_i \\ &\equiv \alpha_0 + \alpha_d D_i + \epsilon_i^*, \end{aligned} \qquad (6)$$

where $\alpha_0 = \beta_0 + \beta_c \gamma_0, \alpha_d = \beta_c \gamma_d$ and $\epsilon_i^* = \beta_c \eta_i + \epsilon_i$. Therefore, the marginal DR model is still linear with a simple parameter conversion, and $\alpha_d = \beta_c \gamma_d$ is the key parameter to present the DR relationship.

## 3.2 Exposure-response models with sigmoid curves

The most commonly used ER models are those representing a sigmoid DR relationship. In statistician's tool box, the logistic model with mean response at a given exposure level $c$ as

$$\mu(c) \equiv E(Y_i | C_i = c) = \frac{\exp(\beta_0 + \beta_c c)}{1 + \exp(\beta_0 + \beta_c c)} \qquad (7)$$

is commonly used for binary or binomial outcomes. We will denote the right-hand side as $\mathrm{expit}(\beta_0 + \beta_c c)$. The same model can be used in the framework of quasi-likelihood or estimating equation to model range-limited outcomes. For example, the survival probability at a given time can be modeled by a quasi logistic regression.

For PK/PD modeling, the Emax model is a classical tool for dose-exposure-response modeling and has been widely used for dose selection for a long time. The 4-parameter Emax ER model is

$$E(Y_i | C_i) = b_0 + \frac{(b_m - b_0)(\exp(C_i)/b_{50})^\gamma}{1 + (\exp(C_i)/b_{50})^\gamma} \qquad (8)$$

where $b_0$ and $b_m$ are the minimum and maximum mean outcome, $b_{50}$, also known as $EC_{50}$, is the exposure level delivering 50% of the maximum effect. Known as the Hill parameter, $\gamma$ determines the slope of increase/decrease of the curve and is often a parameter that is difficult to estimate. For the DR model, we only need to replace $C_i$ with $D_i$ in the model. In our case, with sparsed dose levels, we have made no attempt to accurately estimate $\gamma$, nor is this possible. Thomas [2006] demonstrated that the Emax model is able to model a wide range of increasing monotone curves. The Emax model, apart from the two parameters that determine the range of $Y_i$, can be converted to a logistic model with a reparameterization transformation of $C_i$ [An et al., 2019]. Therefore, the simple logistic model captures the essential sigmoid shape of the Emax model. This 4-parameter Emax model can be made more flexible by adding one more parameter; see Gottschalk and Dunn [2005]. However, we will not consider it, since a dose-finding trial rarely has more than 4 fixed dose levels.

One problem of using the logistic model (and most nonlinear models) in comparison between the DR and DER modeling approaches is due to its non-collapsibility, that is, if $E(Y_i|C_i)$ follows the logistic model (7), $E(Y_i|D_i)$ does not, since $E(\text{expit}(\beta_0 + \beta_c(\gamma D_i + \eta_i)))$ is generally not an expit function anymore. Therefore, not only $\beta_c \gamma \neq \beta_d$ in a logistic DR model with $E(Y_i|D_i) = \text{expit}(\beta_0 + \beta_d D_i)$, but also the parameters of both models cannot be made comparable.

A less used, but very similar to the logistic one, is the probit model. Although rather different mathematically, the model has a very similar curve to the logistic one, hence is sometimes used as an alternative to the logistic model. The probit model defined below has the advantage that a marginal probit model averaging over a normally distributed random variable is still a probit model. To formally define the probit ER model, let the binary response be $Y_i = I(\beta_0 + \beta_c C_i + \epsilon_i > 0)$, with $\epsilon_i \sim N(0,1)$, but is correlated with $\eta_i$. For the marginal DR model to be a probit model, we need a normally distributed zero-mean $\eta_i$ in the DE model. Then $Y_i$ follows a Bernoulli distribution with mean

$$
\begin{aligned}
E(Y_i|D_i) &= P(I(\beta_0 + \beta_c \gamma_d D_i + \beta_c \eta_i + \epsilon_i > 0)) \\
&= \Phi((\beta_0 + \beta_c \gamma_d D_i)\text{var}(\beta_c \eta_i + \epsilon_i)^{-1/2}) \\
&= \Phi(\alpha_0 + \alpha_d D_i)
\end{aligned}
\tag{9}
$$

where the second equation follows from normalizing the random term $\beta_c \eta_i + \epsilon_i$ to a standard normal distributed random variable. Therefore, the marginal DR model follows a probit distribution with parameters

$$(\alpha_0, \alpha_d) = (\beta_0, \beta_c \gamma_d)\text{var}(\beta_c \eta_i + \epsilon_i)^{-1/2}. \tag{10}$$

In the next section, we will consider the DER approach in which we adjust for an estimate of $\eta_i$. The normalization approach will also be used.

## 4 Randomization as IV and control function adjustment

When $\epsilon_i$ and $\eta_i$ are dependent, even for the simple linear ER model, fitting it by least squares (LS) is not valid. Taking the linear models (2) for centered data (hence $\beta_0 = 0$) as an example, the estimate of $\beta_c$ can be written as

$$\hat{\beta}_c = \frac{\sum_{i=1}^n C_i Y_i}{\sum_{i=1}^n C_i^2} = \beta_c + \frac{n^{-1}\sum_{i=1}^n C_i \epsilon_i}{n^{-1}\sum_{i=1}^n C_i^2}$$

where the second term is the confounding bias since the denominator tends to a constant $S_c$ and, since $D_i$ is randomized, the numerator tends to $E(\epsilon_i \eta_i) = \text{cov}(\epsilon_i, \eta_i)$. Therefore, the correlation between $\eta_i$ and $\epsilon_i$ causes a confounding bias. Our goal is to eliminate confounding bias asymptotically, that is, to make $\hat{\beta}_c$ converging to $\beta_c$ when $n \to \infty$.

An approach to eliminating confounding bias is to use randomized doses as an IV. Three key requirements for an IV are: 1) it is correlated with $C_i$, and 2) it does not affect $Y_i$ except through its potential effect on $C_i$, and 3) it does not share common causes with $Y_i$. These requirements are commonly satisfied by randomized dose in dose finding trials, although 2) may not hold exactly for some drugs. CF is a powerful approach based on IV. A CF is a random variable that makes $C_i$ and $\epsilon_i$ independent by conditioning on the CF. Details of IV approaches can be found in textbooks such as Cameron and Trivedi [2005]; Hernan and Robins [2020].

We assume that the DE model is linear as in (12), but the exposure $C_i$ is defined as log-concentration. The use of log-concentration fits the range of linear and probit ER models. But in some ER models using log-concentration may not be appropriate. The work we report here focuses on a few models that allow for comparison between the DR and DER modeling approaches. The key step is to project $\epsilon_i$ to $\eta_i$, hence we assume that $\epsilon_i = E(\epsilon_i|\eta_i) + e_i$, so that we separate the error term in the ER model into two components: one induces confounding but can be controlled by $\eta_i$ or its estimate in the ER model, and the other is a pure random term. This separation suggests predicting $\eta_i$ as the CF. See the algorithm below for how to construct it.

In general, the CF approach consists of two steps:

1. Fit the linear DE model in (2) and obtain the residuals $\hat{\eta}_i = C_i - \hat{\gamma}_0 + \hat{\gamma}_d D_i$.

2. Fit the ER model with $C_i$ and $\hat{\eta}_i$ as covariates.

To see how the CF approach works, note that the problem of fitting the original ER model without adjustment is due to the correlation between $C_i$ and $\epsilon_i$, while the ER model with $\hat{\eta}_i$ as a covariate has controlled the correlated component and the random term in this model reduces to $e_i$. Under some technical conditions, it has been shown that $\hat{\eta}_i \to \eta_i$ weakly when $n \to \infty$ (1) [Imbens and Newey, 2009], [Zhang et al., 2024]. The assumption of the linear DE model can be relaxed, although some assumptions are still required to construct a CF with a general DE model in (1) [Imbens and Newey, 2009], Zhang et al. [2024], but here we will focus on the simple linear DE model. In addition, $E(\epsilon_i|\eta_i)$ can be a nonlinear function such as spline functions of $\eta_i$ to fit the ER model [Zhang et al., 2024].

There are other IV based approaches such as the two-step least squares (2SLS) approach which is the same as the CF one, except in Step 2, a predicted $C_i$ based on the DE model in Step 1 is fitted without $\hat{\eta}_i$ as a covariate. Another approach uses estimating equations with $D_i$ as an IV in it. For linear models such as (12), all approaches are equivalent. For other models, the CF approach is often more efficient but less robust. See, e.g., Cameron and Trivedi [2005] for a detailed introduction to IV based methods. We will focus on the CF approach due to the small sample size in our context.

The 2SLS approach has an important connection to the sequential modeling approach for popPK/PD modeling, since both approaches use the DE model to predict exposure and use the predicted exposure in the ER model. To eliminate any potential bias, the 2SLS approach predicts the mean exposure with dose alone; therefore, all patients with the same dose will have the same predicted exposure. In this way there is no confounding effect in the predicted exposure, but much of the advantage of the DER modeling approach has been lost, e.g., a simple ER nonlinear model may not be identified. Therefore, its use in our setting is limited. In contrast, popPK modeling often uses covariates to predict individual exposure so that rich exposure data can be used to fit ER models. The predicted exposure may still induce confounding, but as covariates are observed, they can be adjusted for in the ER model. This approach would require careful model selection and fitting since the adjustment depends on the correct model specification. If the popPK model also uses random effects in the prediction, unobserved confounding effects may also be included in the predicted exposure. The CF approach avoids the problem of 2SLS approach, but it needs additional assumptions to construct it.

For nonlinear ER models such as the probit model, the same two-step CF approach described in the paragraph before the last one applies. Step 1 is the same as above, and in Step 2, a model

$$E(Y_i|C_i, \hat{\eta}_i) = \Phi(\beta_0^* + \beta_c^* C_i + \beta_\eta^* \hat{\eta}_i) \tag{11}$$

is fitted. In the model, we denote the parameters with a "*" since they are not the same as in the definition of $Y_i = I(\beta_0 + \beta_c C_i + \epsilon_i > 0)$; Wooldridge [2015] gives the details of this approach. However, to use the fitted model to predict the mean response at a given dose, a parameter conversion between the fitted DE and ER models and the marginal DR model is derived in the next section.

# 5 The benefit of dose-exposure-response modeling approaches

An early work to quantify the benefit of using DER modeling for dose finding is Hsu [2009], who demonstrated by simulation the benefit of using DER modeling and quantified the impact of measurement errors on it. In a very simple scenario where both the DE and ER models are all linear for illustration, (Wang [2015], Section 9.2) derived the bound of benefit and the impact of the model parameters in an analytical form. This section examines further this topic, particularly the risk due to confounding factors and differential measurement error, and possible remedies for them.

## 5.1 Linear ER models

We first consider simple linear DE and ER models

$$\begin{aligned} Y_i &= \beta_c C_i + \epsilon_i \\ C_i &= \gamma_d D_i + \eta_i \end{aligned} \tag{12}$$

with centered data, that is, the means of $Y_i, C_i$ and $D_i$ are zero, so that there is no intercept in the models; $\epsilon_i$ and $\eta_i$ are independent zero mean random variables with variances $\mathrm{var}(\epsilon_i) = \sigma_\epsilon^2$ and $\mathrm{var}(\eta_i) = \sigma_\eta^2$, hence $C_i$ is not confounded with $\epsilon_i$. This setting leads to a marginal DR model

$$Y_i = \beta_c \gamma_d D_i + \beta_c \eta_i + \epsilon_i \equiv \alpha_d D_i + \epsilon_i^* \tag{13}$$

where $\alpha_d$ can be estimated by fitting the DR model (13) by LS to $Y_i$ and $D_i$ only, or by $\hat{\gamma}_d \hat{\beta}_c$, where $\hat{\gamma}_d$ and $\hat{\beta}_c$ are the estimates by fitting the DE and ER models separately. An interesting question is: Is the DER modeling more efficient, if so, by how much? Wang [2015] (Section 9.2) shows that the asymptotic variances of the latter can be written as

$$\begin{aligned} \mathrm{var}(\hat{\gamma}_d \hat{\beta}_c) &= (E(\hat{\beta}_c))^2 \mathrm{var}(\hat{\gamma}_d) + (E(\hat{\gamma}_d))^2 \mathrm{var}(\hat{\beta}_c) \\ &\approx (\beta_c^2 \sigma_\eta^2/\sigma_d^2 + \gamma_d^2 \sigma_\epsilon^2/\sigma_c^2)/n \end{aligned} \tag{14}$$

where $\sigma_d^2 = \text{var}(D_i)$ and $\sigma_c^2 = \text{var}(C_i) = \gamma_d^2\sigma_d^2 + \sigma_\eta^2$. The variance of $\hat{\alpha}_d$ is $\text{var}(\hat{\alpha}_d) = n^{-1}(\beta_c^2\sigma_\eta^2 + \sigma_\epsilon^2)/\sigma_d^2$. Then the ratio of the variance of the two estimators is

$$\frac{\text{var}(\hat{\gamma}_d\hat{\beta}_c)}{\text{var}(\hat{\alpha}_d)} \approx \frac{\beta_c^2\sigma_\eta^2/\sigma_d^2 + \gamma_d^2\sigma_\epsilon^2/\sigma_c^2}{(\beta_c^2\sigma_\eta^2 + \sigma_\epsilon^2)/\sigma_d^2}$$
$$= 1 - \frac{\sigma_\eta^2\sigma_\epsilon^2}{(\beta_c^2\sigma_\eta^2 + \sigma_\epsilon^2)(\gamma_d^2\sigma_d^2 + \sigma_\eta^2)} \tag{15}$$

As the second term of (15) is positive, the DER estimator is always more efficient than the DR model one. In addition, the second term tends to zero when $\sigma_\eta^2 \to 0$ or $\sigma_\eta^2 \to \infty$. The maximum gain occurs in between the extremes. This result is based on independent $\epsilon_i$ and $\eta_i$, i.e., no confounding in the ER model hence it can be fitted without any adjustment.

Next, we examine the relative efficiency of DER estimator using the CF approach. Let $\tilde{\beta}_c$ be the estimate of fitting model $Y_i = \beta_c C_i + \beta_\eta \hat{\eta}_i + \epsilon_i$. We compare the variance ratio of the DER estimator $\tilde{\beta}_c\hat{\gamma}_d$ to that of $\hat{\alpha}_d$, the estimated parameter in the linear DR model using only data $D_i$ and $Y_i$. For the variance of $\tilde{\beta}_c$, we take a shortcut using the equivalence between the CF and the 2SLS estimators [Petrin and Kim, 2011]. Since the asymptotic variance of the 2SLS estimator is

$$\text{var}(\tilde{\beta}_c) \approx n^{-1}\sigma_\epsilon^2\text{cov}(D_i, C_i)^{-1}\text{var}(D_i)\text{cov}(D_i, C_i)^{-1}. \tag{16}$$

so is the variance of the CF estimator. Also, with the DE model in (12), we have $\text{cov}(D_i, C_i) = \gamma_d\text{var}(D_i) = \gamma_d\sigma_d^2$, hence $\text{var}(\tilde{\beta}_c) \approx n^{-1}\sigma_\epsilon^2/(\sigma_d^2\gamma_d^2)$. Taking it to (14) to obtain $\text{var}(\hat{\gamma}_d\tilde{\beta}_c)$ and replacing the numerator in (15), we have

$$\frac{\text{var}(\hat{\gamma}_d\tilde{\beta}_c)}{\text{var}(\hat{\alpha}_d)} \approx \frac{\beta_c^2\sigma_\eta^2/\sigma_d^2 + \sigma_\epsilon^2/\sigma_d^2}{(\beta_c^2\sigma_\eta^2 + \sigma_\epsilon^2)/\sigma_d^2} = 1 \tag{17}$$

That is, with linear models (12), there is no gain (and no loss) in using DER if the IV adjustment is needed to adjust for potential confounders. Note that although we only show equal asymptotic variance, the two estimators in the linear model situation are, in fact, numerically identical.

## 5.2 Probit ER models

We derive the marginal DR model via DER modeling approach using probit models. We can find a marginal probit model that combines the DE and ER models and is equivalent to the probit DR model (9) and equivalent parameters. Consequently, we can compare the accuracy of the equivalent parameter estimation by the DER and DR approaches. For simplicity, we will not consider the case of no confounding bias, as we did for the linear model.

With normally distributed $\eta_i$ and $\epsilon_i$, we can write $\epsilon_i = \beta_\eta\eta_i + \xi_i$, where the first term is the part predicted by $\eta_i$ and $\xi_i$ is the residual part independent of $\eta_i$. In particular, denote $\text{cor}(\eta_i, \epsilon_i) = \rho$ and note that $\text{var}(\epsilon_i) = 1$ (the assumption we used for deriving (9)), we have

$$\beta_\eta = \frac{\text{cov}(\eta_i, \epsilon_i)}{\sigma_\eta^2} = \frac{\text{cor}(\eta_i, \epsilon_i)\sigma_\eta}{\sigma_\eta^2} = \rho/\sigma_\eta \tag{18}$$

and $\text{var}(\xi_i) = 1 - \rho^2$ is the residual variance.

The CF approach fits a probit ER model

$$E(Y_i|C_i, \hat{\eta}_i) = \Phi(\beta_0^* + \beta_c^* C_i + \beta_\eta^* \hat{\eta}_i) \qquad (19)$$

where the right-hand side is equivalent to

$$P(\beta_0 + \beta_c C_i + \beta_\eta \hat{\eta}_i + \xi_i \geq 0) \qquad (20)$$

since $P(\beta_0 + \beta_c C_i + \beta_\eta \hat{\eta}_i + \xi_i \geq 0) = P(\beta_0^* + \beta_c^* C_i + \beta_\eta^* \hat{\eta}_i + \xi_i/\sqrt{1-\rho^2} \geq 0)$, where we replace $\eta_i$ with $\hat{\eta}_i$ since the latter weakly converges to the former and the replacement does not affect our comparison. The conversion between the two sets of parameters $\boldsymbol{\beta}^* \equiv (\beta_0^*, \beta_c^*)$ and $\boldsymbol{\beta} \equiv (\beta_0, \beta_c)$ is $\boldsymbol{\beta}^* = \boldsymbol{\beta}(1-\rho^2)^{-1/2}$. For conversion, $\rho$ can be estimated using the following relationship:

$$\beta_\eta^* = \beta_\eta(1-\rho^2)^{-1/2} = \frac{\rho}{\sigma_\eta(1-\rho^2)^{1/2}} \qquad (21)$$

which leads to

$$\rho^2 = \frac{\beta_\eta^* \sigma_\eta^2}{1 + \beta_\eta^* \sigma_\eta^2} \qquad (22)$$

Finally, we replace $C_i$ in the ER model with that in the DE model to the equivalent form of the ER model (20), then the marginal DR model can be written as

$$\begin{aligned} E(Y_i|D_i) &= P(\beta_0 + \beta_c(\gamma_0 + \gamma_d D_i + \eta_i) + \beta_\eta \hat{\eta}_i + \xi_i \geq 0) \\ &= P(\beta_0 + \beta_c \gamma_0 + \beta_c \gamma_d D_i + (\beta_c + \beta_\eta)\hat{\eta}_i + \xi_i \geq 0) \\ &= \Phi\left(\frac{\beta_0 + \beta_c \gamma_0 + \beta_c \gamma_d D_i}{\sqrt{1-\rho^2 + (\beta_c + \beta_\eta)^2 \sigma_\eta^2}}\right) \end{aligned} \qquad (23)$$

Although we only know $\boldsymbol{\beta}^*$, the parameters in (23) can be calculated as $\boldsymbol{\beta} = \boldsymbol{\beta}^*(1-\rho^2)^{1/2}$. Comparing the above with the probit DR model, we find the relationship between the two sets of parameters:

$$(\alpha_0, \alpha_d) = (\beta_0 + \beta_c \gamma_0, \beta_c \gamma_d)(1 - \rho^2 + (\beta_c + \beta_\eta)\sigma_\eta^2)^{-1/2} \qquad (24)$$

Inference based on IV approaches can be based on asymptotic normality of the estimators, which is well developed [Cameron and Trivedi, 2005]. In practice, one can also use bootstrap, which is often more robust than the asymptotic property based ones, and may also be easier for implementation.

Although, in principle, we can compare DER and DR estimators based on the asymptotic properties of the estimated parameters in both models and the relationship between them (24), we prefer to use simulation for comparisons of small-sample properties.

### 5.3 Response estimation using DER models with control function adjustment

In this section, we consider the estimation of responses for a given dose, which is the key use of DR and DER modeling for dose finding and optimization. After fitting a DR model, either directly or via the DER approach, it is easy to estimate the mean response for a given dose. For example, if a probit DR model is fitted, we can estimate $\mu(d) = E(Y_i|D_i = d)$ by

$$\hat{\mu}(d) = \Phi(\hat{\beta}_{d0} + \hat{\beta}_d d) \tag{25}$$

while using the DER model (23) follows the same way. Although there are more parameters, they can be estimated. For DER models such as the probit one, it is more appropriate to use the bootstrap approach for inference, due to the complexity of the marginal DR relationship with several parameters involved.

The model-based approach using the probit model is sensitive to model misspecification. An alternative is a semiparametric approach using empirical means to replace the analytical formulae in the last section. After fitting the probit ER model, $\mu(d)$ can be estimated by

$$\hat{\mu}_{pb}(d) = n^{-1} \sum_{i=1}^{n} \Phi(\hat{\beta}_0^* + \hat{\beta}_c^* \hat{C}_i(d) + \hat{\beta}_v^* \hat{\eta}_i) \tag{26}$$

where $\hat{C}_i(d) = \hat{\gamma}_0 + \hat{\gamma}_d d + \hat{\eta}_i$. This estimate still needs a normally distributed $e_i$, but not such an $\eta_i$.

The same applies to the logistic regression ER model

$$E(Y|C_i, \eta_i) = \text{expit}(\beta_0 + \beta_c C_i + \beta_\eta \eta_i)) \tag{27}$$

Although a constant marginal odds ratio may not exist, the logistic model-based estimator for the mean response at a given dose $d$

$$\hat{\mu}_{lg}(d) = n^{-1} \sum_{i=1}^{n} \text{expit}(\hat{\beta}_0^* + \hat{\beta}_c^* \hat{C}_i(d) + \hat{\beta}_v^* \hat{\eta}_i) \tag{28}$$

is valid. Again, in the next section we will use simulation for comparisons of the small sample properties of the response estimation approaches.

# 6 A simulation study

To examine the benefit of using the combined DER estimator, compared to using a DR model directly, we conduct a simulation study to compare the two approaches and focus on probit models so that a fair comparison can be made between them. Results using a logistic model for the empirical mean approach to response estimation are also presented. Simulation for linear ER models is not needed due to the analytical result in Section 5.

For simulation, we generate $C_i$ using a linear DE model and $Y_i$ using a probit model with different levels of confounding effect and sample sizes. We set $\beta_0 = -3, \beta_c = 1, \gamma_0 = 0, \gamma_d = 1$, $\sigma_\epsilon = 1, \sigma_\eta = 1$, with $\rho$ varying from 0 to 0.9 and the sample size from 40 to 120. Note that $C_i$ and $D_i$ represent PK levels and doses at steady state on the log scale in the motivating scenario. This model setting gives $(\alpha_0, \alpha_d) = (-2.12, 0.71)$ when $\rho = 0$. Two scenarios are considered:

1. Scenario 1: Dose levels are (1,2,3,4,5) units. This dose range gives a symmetric pattern $E(Y_i|d) = (0.08, 0.24, 0.50, 0.76, 0.92)$ when $\rho = 0$.

2. Scenario 2: The dose levels are (1,2,3,4,5)/1.5 unit. This dose range covers the lower part of the sigmoid curve and gives $E(Y_i|d) = (0.05, 0.12, 0.24, 0.40, 0.60)$ when $\rho = 0$. $\hat{\eta}_i$ are taken from the residuals in the fitted DE model.

For each scenario, 10000 simulation runs are performed. To examine efficiency, we compare the bias, variance, and DER estimates with and without CF adjustment, and the variance and MSE ratios of the DER estimates with those of the DR estimates. The R code for the simulation study is given in the Appendix.

The results of Scenario 1 with varying sample size and $\rho$ are summarized in Table 1. Both the DR and adjusted DER estimates have moderate biases, especially when $n = 40$, which are likely due to nonlinearity, since at least the DR estimate is not affected by confounding, due to the randomized dose. Interestingly, in general, the bias of the adjusted DER estimator is lower than that of the DR one. The variance ratios are all less than one, which shows the benefit of DER approaches in general, although in some situations the efficiency gain is small. The benefit is higher when sample sizes are small, which is expected, since the extra information in DER modeling may be more important with small sample sizes. The MSE ratio shows a similar trend for adjusted DER estimator. When $\rho = 0$ there is no confounding error such that the DER estimator without CF adjustment is also valid. For all sample sizes, the loss of efficiency due to the CF adjustment is substantial. The biases of DER estimates increase with $\rho$, while the variance ratio shows that the unadjusted DER approach has less variability than the DR and adjusted DER estimates. With moderate confounders ($\rho = 0.3$), the MSE ratios of the unadjusted DER estimates are still lower than those of the adjusted ones, while for severe confounding ($\rho = 0.9$) the MSE of the unadjusted estimate are higher than the DR and adjusted DER estimates. The sample size also plays a role in the comparison, since with $n = 40$, the adjusted and unadjusted DER estimates have almost the same MSE when $\rho = 0.6$, but this is not true for larger sample sizes.

Table 2 gives the results in Scenario 2, which shows very similar patterns as we found in Scenario 1, although the amount of benefit changes moderately. This result suggests the benefit of the DER approach when the fixed dose set only covers a part of the entire dose range, as often occurs in dose finding trials.

The simulated data are also used to predict the DR curve using both the DR and DER approaches. Figure 1 shows the variance ratio of the DER estimator vs. the DR estimator at each dose level, for sample size = 40 and 80 and $\rho = 0, 0.3, 0.6, 0.9$ in Scenario 1. The DER estimators are overall more efficient than DR, with their variance ratios varying with the extent of confounding measured by $\rho$, the sample size $n$, and the methods with CF (and without it, labeled "Unadj." for $\rho = 0$ only, which is comparable with the curve of $\rho = 0$). The results are consistent with those of parameter estimators in the marginal probit model. The DER approach without CF adjustment has a much lower variance ratio, especially at the two ends of the curve. This may be due to the fact that, intuitively, the DR estimator has a lower variance in the middle of the curve than at the two ends. Nevertheless, the DER still shows some benefit. As expected, the estimator with CF adjustment has higher variances than that without. However, unlike the linear ER model case, the DER estimator is still more efficient than the DR estimator. Comparing the performance of the DER estimator with different $\rho$ values, we find that the variance ratio generally reduces with increasing $\rho$, especially around the medium dose. This is also expected, since the adjustment reduces the random component in the ER model. In addition, the CF estimator with $n = 80$ performs better than when $n = 40$, since it is well known that IV-based approaches, including the CF one, generally perform better with larger sample sizes. An interesting observation is that the DER estimator with CF performs better around the medium dose, while the one without CF performs better at the two ends.

Table 1: Summary of biases, variances and MSE, and variances and MSE ratios of DER estimator with and without CF adjustment to the DR estimator using Probit regressions by $\rho$ the proportion of confounded variation in exposure in terms of variances, with sample sizes 40, 80 and 120: Scenario 1.

| n | $\rho$ | CF adj. | Bias (DR) $\alpha_0$ | $\alpha_d$ | Bias (DER) $\alpha_0$ | $\alpha_d$ | Var(DR) $\alpha_0$ | $\alpha_d$ | V(DER)/V(DR) $\alpha_0$ | $\alpha_d$ | MSE(DER)/MSE(DR) $\alpha_0$ | $\alpha_d$ |
|---|---|---|---|---|---|---|---|---|---|---|---|---|
| 40 | 0.0 | No | -0.305 | 0.101 | -0.114 | 0.037 | 1.805 | 0.188 | 0.15 | 0.12 | 0.15 | 0.12 |
| 40 | 0.0 | Yes | -0.305 | 0.101 | -0.227 | 0.075 | 1.805 | 0.188 | 0.66 | 0.63 | 0.65 | 0.63 |
| 40 | 0.3 | No | -0.288 | 0.094 | -0.243 | 0.079 | 1.437 | 0.150 | 0.18 | 0.15 | 0.21 | 0.18 |
| 40 | 0.3 | Yes | -0.288 | 0.094 | -0.212 | 0.069 | 1.437 | 0.150 | 0.57 | 0.56 | 0.57 | 0.56 |
| 40 | 0.6 | No | -0.214 | 0.069 | -0.525 | 0.173 | 0.884 | 0.087 | 0.31 | 0.26 | 0.59 | 0.57 |
| 40 | 0.6 | Yes | -0.214 | 0.069 | -0.154 | 0.049 | 0.884 | 0.087 | 0.59 | 0.58 | 0.59 | 0.58 |
| 40 | 0.9 | No | -0.159 | 0.051 | -0.926 | 0.306 | 0.557 | 0.056 | 0.54 | 0.45 | 1.99 | 2.04 |
| 40 | 0.9 | Yes | -0.159 | 0.051 | -0.094 | 0.029 | 0.557 | 0.056 | 0.60 | 0.59 | 0.59 | 0.58 |
| 80 | 0.0 | No | -0.091 | 0.031 | -0.040 | 0.013 | 0.248 | 0.024 | 0.47 | 0.42 | 0.46 | 0.41 |
| 80 | 0.0 | Yes | -0.091 | 0.031 | -0.070 | 0.023 | 0.248 | 0.024 | 0.92 | 0.91 | 0.91 | 0.90 |
| 80 | 0.3 | No | -0.107 | 0.034 | -0.174 | 0.056 | 0.233 | 0.022 | 0.50 | 0.44 | 0.60 | 0.55 |
| 80 | 0.3 | Yes | -0.107 | 0.034 | -0.087 | 0.027 | 0.233 | 0.022 | 0.91 | 0.91 | 0.90 | 0.90 |
| 80 | 0.6 | No | -0.089 | 0.027 | -0.459 | 0.150 | 0.196 | 0.019 | 0.61 | 0.53 | 1.62 | 1.68 |
| 80 | 0.6 | Yes | -0.089 | 0.027 | -0.067 | 0.020 | 0.196 | 0.019 | 0.89 | 0.89 | 0.87 | 0.88 |
| 80 | 0.9 | No | -0.078 | 0.024 | -0.866 | 0.286 | 0.175 | 0.016 | 0.78 | 0.67 | 4.89 | 5.50 |
| 80 | 0.9 | Yes | -0.078 | 0.024 | -0.051 | 0.015 | 0.175 | 0.016 | 0.80 | 0.80 | 0.79 | 0.79 |
| 120 | 0.0 | No | -0.053 | 0.017 | -0.022 | 0.007 | 0.146 | 0.014 | 0.50 | 0.43 | 0.49 | 0.43 |
| 120 | 0.0 | Yes | -0.053 | 0.017 | -0.039 | 0.013 | 0.146 | 0.014 | 0.93 | 0.93 | 0.92 | 0.92 |
| 120 | 0.3 | No | -0.065 | 0.020 | -0.150 | 0.049 | 0.140 | 0.013 | 0.53 | 0.47 | 0.67 | 0.62 |
| 120 | 0.3 | Yes | -0.065 | 0.020 | -0.052 | 0.016 | 0.140 | 0.013 | 0.92 | 0.93 | 0.91 | 0.92 |
| 120 | 0.6 | No | -0.053 | 0.016 | -0.434 | 0.142 | 0.123 | 0.012 | 0.63 | 0.54 | 2.11 | 2.26 |
| 120 | 0.6 | Yes | -0.053 | 0.016 | -0.039 | 0.011 | 0.123 | 0.012 | 0.90 | 0.90 | 0.89 | 0.90 |
| 120 | 0.9 | No | -0.048 | 0.014 | -0.838 | 0.277 | 0.108 | 0.010 | 0.77 | 0.67 | 7.15 | 8.11 |

Figure 2 shows the variance ratio for the same settings as in Figure 1, except in Scenario 2. Although the symmetric pattern of the curves has changed, some general trend, e.g., the efficiency of the DER estimator with CF generally increases with $\rho$ and the estimator without CF performs better than that with CF, still holds. However, there is no pattern as clear as in Figure 1 among the curves of the DER estimator with CF.

Finally, Figures 3 and 4 show the variance ratio of the DER estimators (with and without CF) using logistic regression with empirical sample means of the expit function (28), for the same settings as in Figures 1 and 2. They show similar patterns of the latter, except that the DER estimators with empirical means are less efficient than the normal distribution-based ones.

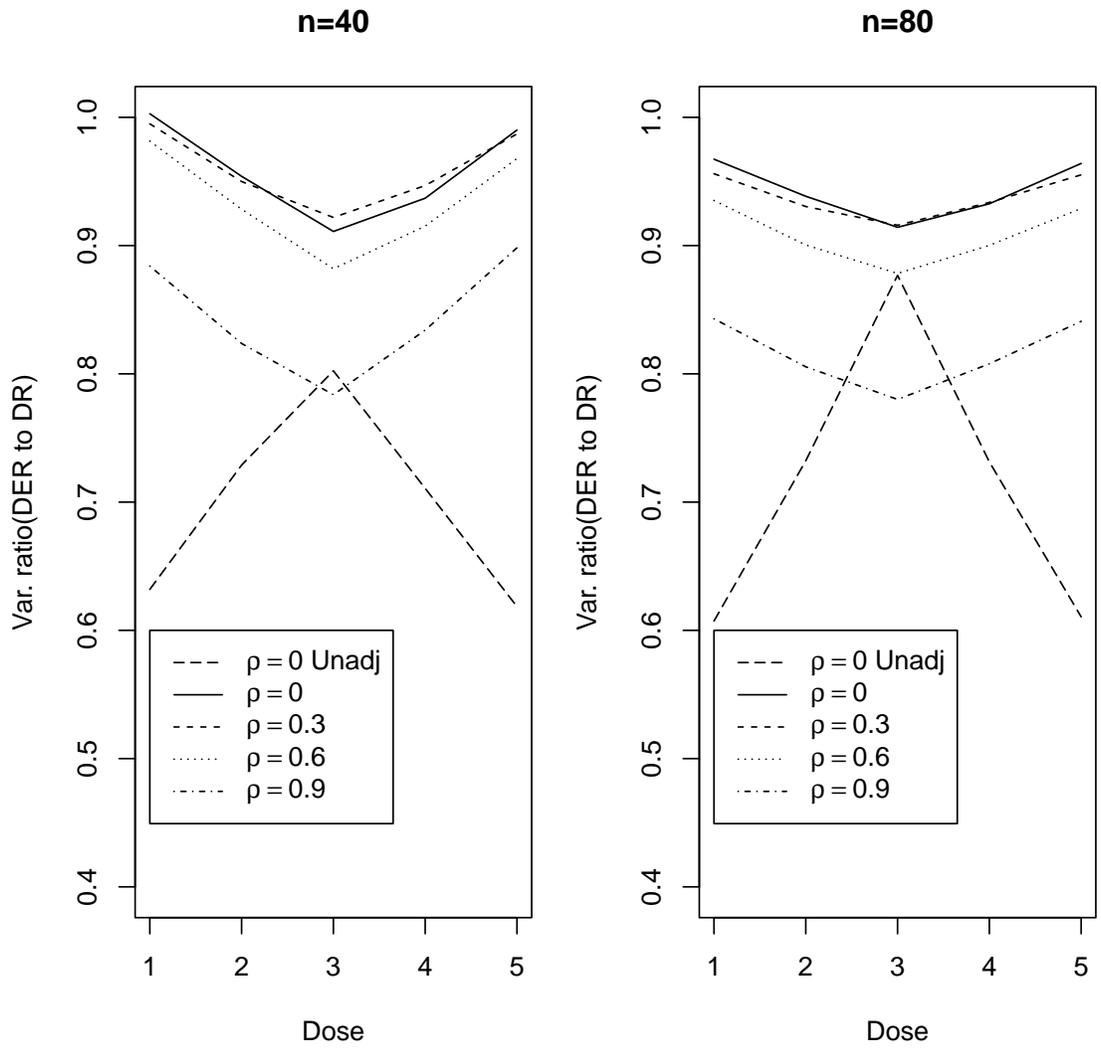

Figure 1: Variance ratio of the DER estimator (with CF and without CF adjustment ($\rho = 0$ only)) to the DR estimator using Probit regressions by $\rho$ the proportion of confounded variation in exposure in terms of variances, with sample sizes 40 and 80: Scenario 1.

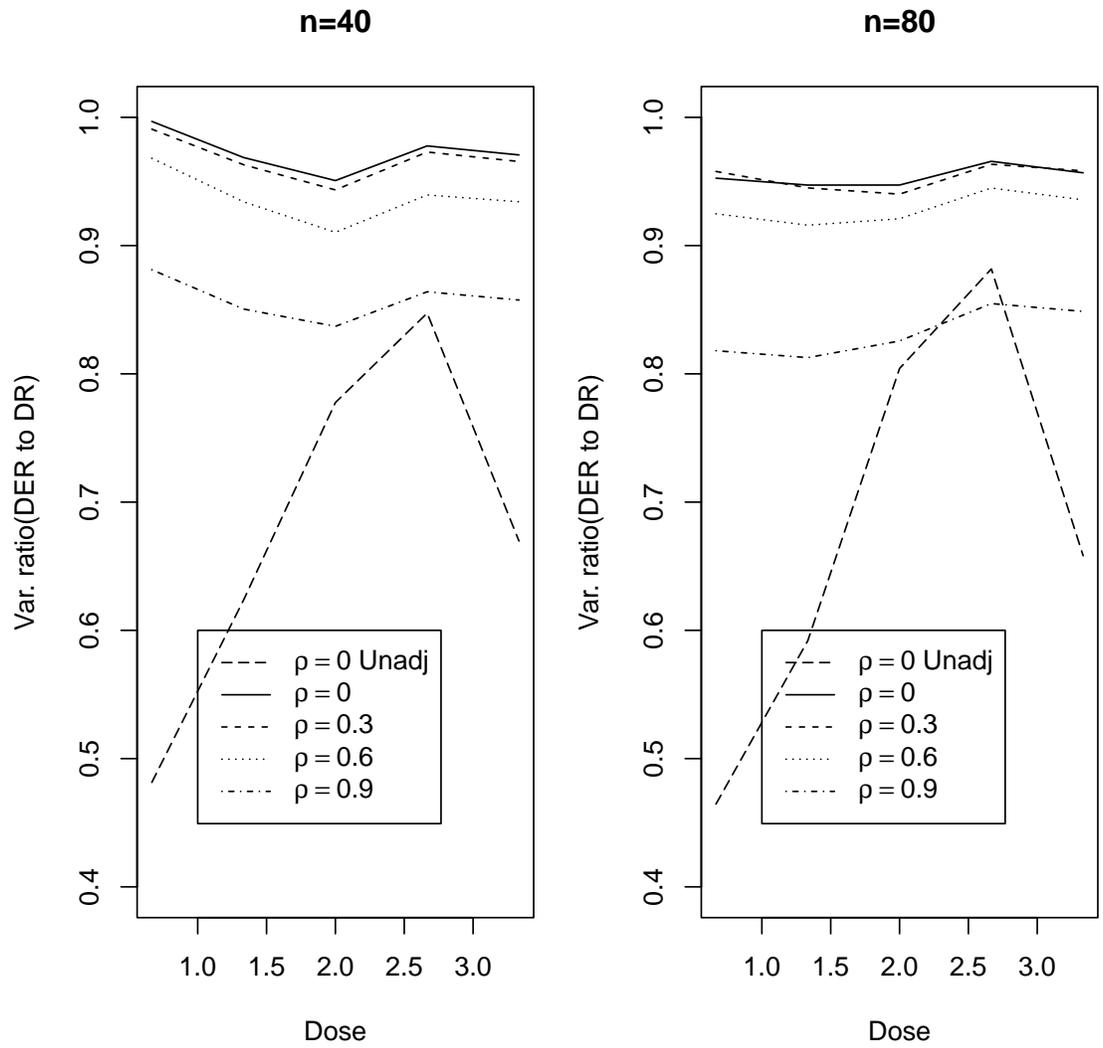

Figure 2: Variance ratio of the DER estimator (with CF and without CF adjustment ($\rho = 0$ only)) to the DR estimator using Probit regressions by $\rho$ the proportion of confounded variation in exposure in terms of variances, with sample sizes 40 and 80: Scenario 2 .

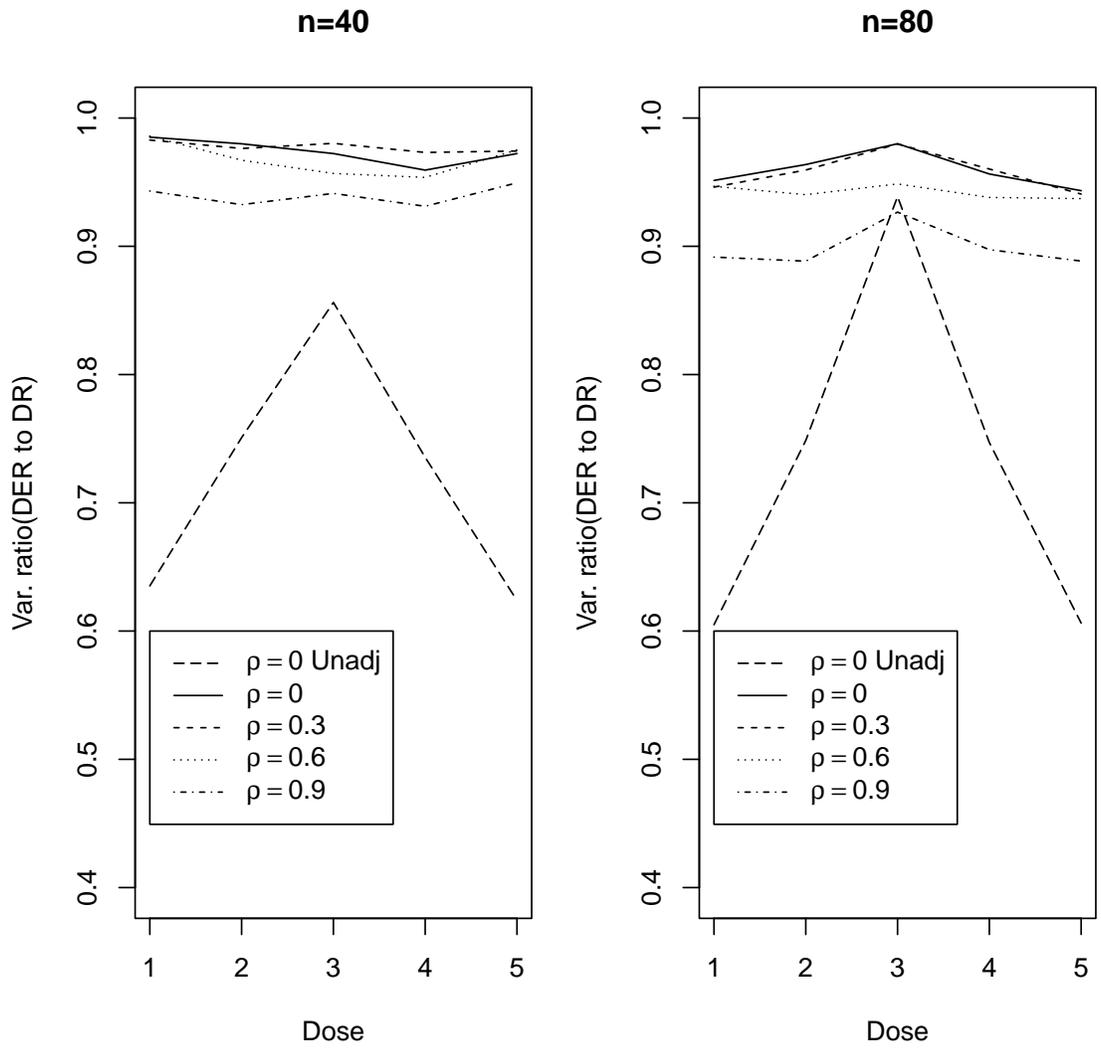

Figure 3: Variance ratio of the DER estimator (with CF and without CF adjustment ($\rho = 0$ only)) to the DR estimator using logistic regressions and empirical means (28) by $\rho$ the proportion of confounded variation in exposure in terms of variances, with sample sizes 40 and 80: Scenario 1

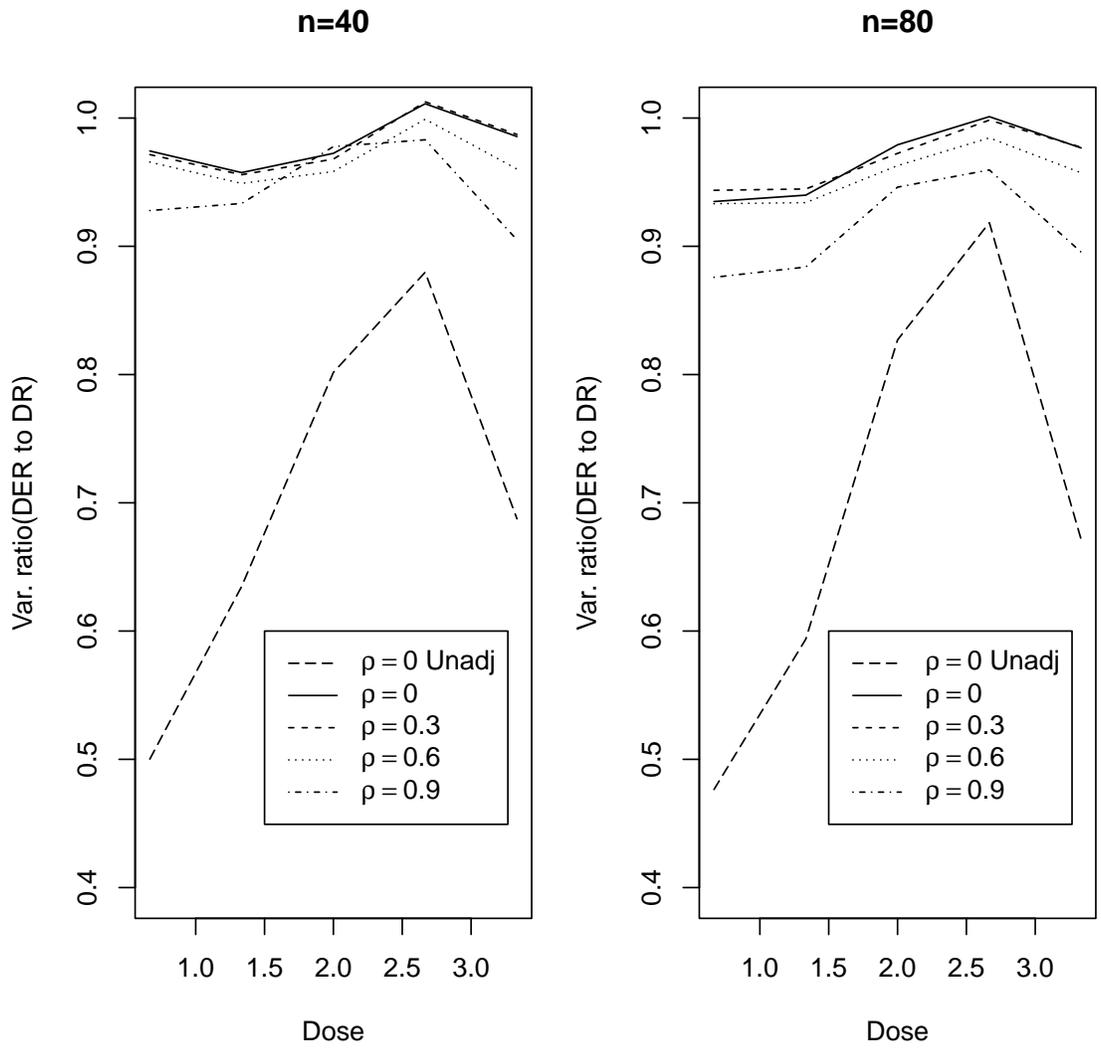

Figure 4: Variance ratio of the DER estimator (with CF and without CF adjustment ($\rho = 0$ only)) to the DR estimator using logistic regressions and empirical means (28) by $\rho$ the proportion of confounded variation in exposure in terms of variances, with sample sizes 40 and 80: Scenario 2.

Table 2: Summary of biases, variances and MSE, and variances and MSE ratios of DER estimators with and without CF adjustment to the DR estimator using Probit regressions by $\rho$: the proportion of confounded variation in exposure in terms of variances, with sample sizes 40, 80 and 120: Scenario 2.

| n | $\rho$ | CF adj. | Bias (DR) $\alpha_0$ | Bias (DR) $\alpha_d$ | Bias (DER) $\alpha_0$ | Bias (DER) $\alpha_d$ | Var(DR) $\alpha_0$ | Var(DR) $\alpha_d$ | V(DER)/V(DR) $\alpha_0$ | V(DER)/V(DR) $\alpha_d$ | MSE(DER)/MSE(DR) $\alpha_0$ | MSE(DER)/MSE(DR) $\alpha_d$ |
|---|---|---|---|---|---|---|---|---|---|---|---|---|
| 40 | 0.0 | No | -0.273 | 0.095 | -0.085 | 0.028 | 1.311 | 0.173 | 0.17 | 0.19 | 0.17 | 0.18 |
| 40 | 0.0 | Yes | -0.273 | 0.095 | -0.212 | 0.073 | 1.311 | 0.173 | 0.71 | 0.75 | 0.71 | 0.74 |
| 40 | 0.3 | No | -0.259 | 0.090 | -0.204 | 0.078 | 1.366 | 0.191 | 0.17 | 0.17 | 0.19 | 0.20 |
| 40 | 0.3 | Yes | -0.259 | 0.090 | -0.193 | 0.066 | 1.366 | 0.191 | 0.59 | 0.63 | 0.59 | 0.63 |
| 40 | 0.6 | No | -0.198 | 0.068 | -0.491 | 0.200 | 0.659 | 0.102 | 0.35 | 0.32 | 0.68 | 0.69 |
| 40 | 0.6 | Yes | -0.198 | 0.068 | -0.144 | 0.048 | 0.659 | 0.102 | 0.76 | 0.79 | 0.74 | 0.77 |
| 40 | 0.9 | No | -0.147 | 0.052 | -0.872 | 0.371 | 0.479 | 0.080 | 0.56 | 0.46 | 2.05 | 2.11 |
| 40 | 0.9 | Yes | -0.147 | 0.052 | -0.087 | 0.029 | 0.479 | 0.080 | 0.71 | 0.74 | 0.69 | 0.73 |
| 80 | 0.0 | No | -0.090 | 0.030 | -0.023 | 0.006 | 0.292 | 0.046 | 0.34 | 0.32 | 0.33 | 0.32 |
| 80 | 0.0 | Yes | -0.090 | 0.030 | -0.071 | 0.023 | 0.292 | 0.046 | 0.90 | 0.91 | 0.89 | 0.91 |
| 80 | 0.3 | No | -0.096 | 0.032 | -0.149 | 0.059 | 0.260 | 0.042 | 0.38 | 0.36 | 0.45 | 0.43 |
| 80 | 0.3 | Yes | -0.096 | 0.032 | -0.076 | 0.025 | 0.260 | 0.042 | 0.89 | 0.91 | 0.88 | 0.90 |
| 80 | 0.6 | No | -0.096 | 0.033 | -0.440 | 0.184 | 0.223 | 0.037 | 0.47 | 0.41 | 1.28 | 1.28 |
| 80 | 0.6 | Yes | -0.096 | 0.033 | -0.076 | 0.026 | 0.223 | 0.037 | 0.87 | 0.88 | 0.86 | 0.88 |
| 80 | 0.9 | No | -0.083 | 0.030 | -0.828 | 0.357 | 0.190 | 0.033 | 0.65 | 0.52 | 4.10 | 4.23 |
| 80 | 0.9 | Yes | -0.083 | 0.030 | -0.057 | 0.020 | 0.190 | 0.033 | 0.77 | 0.79 | 0.76 | 0.78 |
| 120 | 0.0 | No | -0.059 | 0.019 | -0.016 | 0.003 | 0.169 | 0.027 | 0.38 | 0.36 | 0.38 | 0.35 |
| 120 | 0.0 | Yes | -0.059 | 0.019 | -0.044 | 0.013 | 0.169 | 0.027 | 0.91 | 0.92 | 0.90 | 0.92 |
| 120 | 0.3 | No | -0.060 | 0.019 | -0.135 | 0.053 | 0.161 | 0.026 | 0.40 | 0.37 | 0.50 | 0.47 |
| 120 | 0.3 | Yes | -0.060 | 0.019 | -0.046 | 0.014 | 0.161 | 0.026 | 0.90 | 0.91 | 0.89 | 0.91 |
| 120 | 0.6 | No | -0.061 | 0.019 | -0.423 | 0.178 | 0.139 | 0.024 | 0.48 | 0.41 | 1.72 | 1.72 |
| 120 | 0.6 | Yes | -0.061 | 0.019 | -0.047 | 0.014 | 0.139 | 0.024 | 0.86 | 0.88 | 0.86 | 0.87 |
| 120 | 0.9 | No | -0.050 | 0.017 | -0.809 | 0.350 | 0.118 | 0.021 | 0.66 | 0.52 | 6.07 | 6.32 |
| 120 | 0.9 | Yes | -0.050 | 0.017 | -0.037 | 0.012 | 0.118 | 0.021 | 0.76 | 0.78 | 0.76 | 0.77 |

# 7 Discussion

Randomized dose-finding trials are widely used for dose selection, for which the DR relationship is important information. Although DER approaches that include sequential modeling approaches in pharmacometrics have been widely used to predict the DR relationship, a systematic assessment of its benefit, especially with adjustment for unobserved confounding factors, is lacking. We have filled a part of the gap by examining the benefit of using DER modeling with and without adjustment for unobserved confounding factors and present some theoretical and simulation results. Although we have only provided results for linear and probit ER models, the results represent some general scenarios. The probit model is very similar to the logistic model, in particular for prediction purposes, hence our conjecture is that logistic models also behave similarly in the DR and DER modeling. The Emax models have a similar sigmoid curve as a logistic model, but with the range of mean response as additional parameters. The impact of estimating these two parameters on the benefit of DER modeling needs further research. Another type of nonlinear models that are collapsible are the Poisson models for count data or the like. Using CF for adjustment for Poisson models was proposed in Wang [2012]. However, we did not examine them here since these models are less popular than linear and sigmoid models in DER modeling. Also, we have focused on linear DE models, which do not represent a non-linear dose-PK relationship of some drugs. Nevertheless, our approach still applies, as long as a CF can be estimated, e.g., when the DE model can be written as $C_i = h(D_i, \gamma) + \eta_i$ and the model can be correctly identified. This may require more careful model selection and diagnosis, and sensitivity analysis if appropriate.

The model setting of (1) is very general, but an important assumption is that the ER model is an empirical model for *observed* exposure-response relationship. It may be different from a mechanistic model to describe, e.g., the exposure at the effect site and the response. In such a situation, the exposure is observed with error and the measurement error may lead to inconsistency of the mechanistic model parameters. However, since our goal is the estimation of $E(Y_i|d)$, rather than a mechanistic ER model, as long as the empirical DR model is correct (which can be checked), our approach without considering the measurement error is still valid. See Wang [2015] for details on measurement errors in ER modeling.

An interesting topic we have not covered in this work is the comparison between DR modeling and the DER approach using simultaneous modeling based on joint models of DE and ER relationships. Early work [Zhang et al., 2003a,b] investigates the performance of simultaneous modeling, comparing with the sequential modeling used in this work for its best performance in the ideal case, as well as its robustness, but without considering confounding adjustment. Intuitively, the joint modeling approach can be more efficient if the joint model (including the joint distribution of exposure and response) is correct but is less robust, since a misspecified model may have undesirable impact on the fitting of the other one. For this reason, in a different setting but for a technically similar problem, approaches are proposed to isolate the modeling of the ER relationship for robustness [McCandless et al., 2010], which is essentially the sequential modeling approach we used. Nevertheless, a simultaneous modeling approach for the causal DER relationship is a very interesting topic for future research.

In summary, we have evaluated the benefit of using the DER model approach compared to direct DR modeling for DR relationship estimation. In the linear model setting, the DER approach is more efficient than the DR approach without confounding adjustment, but the benefit disappears when confounding adjustment is needed. In the probit model setting and when confounding of the E-R relationship is suspected, a DER-CF approach is beneficial in terms of variance reduction over a DR approach, regardless of whether there is actually confounding or not. The benefit can be substantial for small sample sizes. However, it decreases and becomes 'negligible' when the sample size increases.

# 8 Appendix: R-code for Table 1/2

```
library(xtable)
#  Table 1/2
# The following code reproduce Table 1 (set Scal=1) and 2 (set Scal=1.5)
#adj=1 for CF adjusted, adj=0 for unadjusted
simub=function(ns=40,rho=0,scal=1,nsimu=10000,adj=1){
set.seed(123)
# The true DR model parameters are estimated with very large nsub
nsub=200000
dose=(1:5)/scal    #Dose levels
di=rep(dose, length=nsub)
ui=rnorm(nsub)    # Confounding
av=rho
bv=sqrt(1-av^2)
Ci=di+av*ui+bv*rnorm(nsub)-3   #Exposure data
yi=ifelse(Ci+av*ui+bv*rnorm(nsub)>0,1,0) #Response data
#Estimated "true" alphas in DR model with n=200000
gold=glm(yi~di,family = binomial(link = "probit"))$coef

nsub=ns
Out=NULL
for (j in 1:nsimu){
  di=rep(dose, length=nsub)
  ui=rnorm(nsub)              # Confounding
  Ci=di+av*ui+bv*rnorm(nsub)-3      #Exposure data
  yi=ifelse(Ci+av*ui+bv*rnorm(nsub)>0,1,0) #Response data
  fitdr=glm(yi~di,family = binomial(link = "probit"))
  biasdr=fitdr$coef-gold #Bias of DR model estimates
  fitde=lm(Ci~di)          #DE model
  predc=fitde$coef[1]+fitde$coef[2]*di
  vi=Ci-predc             #hat eta_i
  sigv2=sum(fitde$residuals^2)/(nsub-2)
  if (adj==1){  #CF adjustment
    fiter=glm(yi~Ci+vi,family = binomial(link = "probit"))$coef #Model Eq20
    rho2=(fiter[3]^2*sigv2)/(1+fiter[3]^2*sigv2)   #Eq 22
    fiter=fiter*sqrt(1-rho2)
    #Eq 24
    beta=c(fiter[1]+fiter[2]*fitde$coef[1],fiter[2]*fitde$coef[2])/sqrt(1+(fiter[2]+fiter[3])^2*sigv2-
  }
  else {  #No CF adjustment
    fiter=glm(yi~Ci,family = binomial(link = "probit"))$coef
    beta=c(fiter[1]+fiter[2]*fitde$coef[1],fiter[2]*fitde$coef[2])/sqrt(1+(fiter[2])^2*sigv2)
  }
  biasder=beta-gold   #Bias
  Out=rbind(Out, c(biasdr,biasder))
}
bias=apply(Out,2,mean, na.rm=T)
jk=apply(Out,2,var,na.rm=T)
c(ns,rho,adj,bias, jk[1:2],jk[3:4]/jk[1:2])
}
```

```r
Pout=NULL
Pout=simub(adj=0)
Pout=rbind(Pout,simub())
Pout=rbind(Pout,simub(rho=0.3))
Pout=rbind(Pout,simub(rho=0.6))
Pout=rbind(Pout,simub(rho=0.9))

Pout=rbind(Pout,simub(ns=80,adj=0))
Pout=rbind(Pout,simub(ns=80))
Pout=rbind(Pout,simub(ns=80,rho=0.3))
Pout=rbind(Pout,simub(ns=80,rho=0.6))
Pout=rbind(Pout,simub(ns=80,rho=0.9))

Pout=rbind(Pout,simub(ns=120,adj=0))
Pout=rbind(Pout,simub(ns=120))
Pout=rbind(Pout,simub(ns=120,rho=0.3))
Pout=rbind(Pout,simub(ns=120,rho=0.6))
Pout=rbind(Pout,simub(ns=120,rho=0.9))

xtable(Pout,digits=c(0,0,1,0,rep(3,8)))

#  Figures
#Probit model method to reproduce Figures 1 (scal=1) and 2 (scal=1.5)
simu=function(ns=40,rho=0,scal=1.5,nsimu=10000,adj=1){
  set.seed(123)
  nsub=200000
  dose=(1:5)/scal
  di=rep(dose, length=nsub)
  ui=rnorm(nsub)
  av=rho
  bv=sqrt(1-av^2)
  Ci=di+av*ui+bv*rnorm(nsub)-3
  yi=ifelse(Ci+av*ui+bv*rnorm(nsub)>0,1,0)
  gold=lm(yi~as.factor(di)-1)$coef

  nsub=ns
  Out=NULL
  for (j in 1:nsimu){
    di=rep(dose, length=nsub)
    ui=rnorm(nsub)
    Ci=di+av*ui+bv*rnorm(nsub)-3
    yi=ifelse(Ci+av*ui+bv*rnorm(nsub)>0,1,0)
    fitdr=glm(yi~di,family = binomial(link = "probit")) #DR model
    pred=pnorm(fitdr$coef[1]+fitdr$coef[2]*dose)-gold #Bias of DR prediction
    fitde=lm(Ci~di)                                 #DE model
    predc=fitde$coef[1]+fitde$coef[2]*dose
    vi=Ci-(fitde$coef[1]+fitde$coef[2]*di)          #hat eta_i
    sigv2=sum(fitde$residuals^2)/(nsub-2)
    if (adj==1){
      fiter=glm(yi~Ci+vi,family = binomial(link = "probit"))$coef
```

```r
      rho2=(fiter[3]^2*sigv2)/(1+fiter[3]^2*sigv2)   # Eq 22
      fiter=fiter*sqrt(1-rho2)
      lpd=(fiter[1]+fiter[2]*predc)/sqrt(1+(fiter[2]+fiter[3])^2*sigv2-rho2) #Eq 23
    }
    else {
      fiter=glm(yi~Ci,family = binomial(link = "probit"))$coef
      lpd=(fiter[1]+fiter[2]*predc)/sqrt(1+fiter[2]^2*sigv2)
    }
    predder=pnorm(lpd)-gold
    Out=rbind(Out, c(pred,predder))
  }
  jk=apply(Out,2,var,na.rm=T)
  jk[6:10]/jk[1:5]
}

Pout=NULL
Pout=simu(adj=0)
Pout=rbind(Pout,simu())
Pout=rbind(Pout,simu(rho=0.3))
Pout=rbind(Pout,simu(rho=0.6))
Pout=rbind(Pout,simu(rho=0.9))

Pout=rbind(Pout,simu(ns=80,adj=0))
Pout=rbind(Pout,simu(ns=80))
Pout=rbind(Pout,simu(ns=80,rho=0.3))
Pout=rbind(Pout,simu(ns=80,rho=0.6))
Pout=rbind(Pout,simu(ns=80,rho=0.9))
scal=1.5
dose=(1:5)/scal

pdf(file="PDR15.pdf")
par(mar=c(5.1, 4.1, 4.1, 1.1),mfrow=c(1,2))
plot(dose,ylim=c(0.4,1),Pout[1,],xlab="Dose",ylab="Var. ratio(DER to DR)",type="l",main="n=40",lty=5)
lines(dose,ylim=c(0,1),Pout[2,],type="l",lty=1)
lines(dose,ylim=c(0,1),Pout[3,],type="l",lty=2)
lines(dose,ylim=c(0,1),Pout[4,],type="l",lty=3)
lines(dose,ylim=c(0,1),Pout[5,],type="l",lty=4)
legend(1,0.6,legend=c(expression(paste(rho == 0, " Unadj")),
                      expression(rho == 0),expression(rho == 0.3),expression(rho == 0.6),expression(rh

plot(dose,ylim=c(0.4,1),Pout[6,],xlab="Dose",ylab="Var. ratio(DER to DR)",type="l",main="n=80",lty=5)
lines(dose,ylim=c(0,1),Pout[7,],type="l",lty=1)
lines(dose,ylim=c(0,1),Pout[8,],type="l",lty=2)
lines(dose,ylim=c(0,1),Pout[9,],type="l",lty=3)
lines(dose,ylim=c(0,1),Pout[10,],type="l",lty=4)
legend(1,0.6,legend=c(expression(paste(rho == 0, " Unadj")),
                      expression(rho == 0),expression(rho == 0.3),expression(rho == 0.6),expression(rh
dev.off()

# Empirical mean method to reproduce Figures 3 (scal=1) and 4 (scal=1.5)
```

```r
simu=function(ns=40,rho=0,scal=1.5,nsimu=10000,adj=1){
  set.seed(123)
  nsub=200000
  dose=(1:5)/scal
  di=rep(dose, length=nsub)
  ui=rnorm(nsub)
  av=rho
  bv=sqrt(1-av^2)
  Ci=di+av*ui+bv*rnorm(nsub)-3
  yi=ifelse(Ci+av*ui+bv*rnorm(nsub)>0,1,0)
  # Estimated "true" E(Y|D)
  gold=lm(yi~as.factor(di)-1)$coef

  nsub=ns
  Out=NULL
  for (j in 1:nsimu){
    di=rep(dose, length=nsub)
    ui=rnorm(nsub)
    Ci=di+av*ui+bv*rnorm(nsub)-3
    yi=ifelse(Ci+av*ui+bv*rnorm(nsub)>0,1,0)
    fitdr=glm(yi~di,family = binomial(link = "probit")) #DR model
    pred=pnorm(fitdr$coef[1]+fitdr$coef[2]*dose)-gold #DR prediction
    fitde=lm(Ci~di)                                 #DE model
    predc=fitde$coef[1]+fitde$coef[2]*dose
    ldose=rep(dose,rep(nsub,length(dose)))    #Each dose for all subjects
    vi=Ci-(fitde$coef[1]+fitde$coef[2]*di)    #hat eta_i
    lvi=rep(vi,length(dose))                  #hat eta_i for each dose
    lpredc=fitde$coef[1]+fitde$coef[2]*ldose+lvi #predicted Ci(d) for all doses
    if (adj==1){ #CF adjustment
      fiter=glm(yi~Ci+vi,family = binomial(link = "logit"))$coef
      lpd=fiter[1]+fiter[2]*lpredc+fiter[3]*lvi
      ppred=1/(1+exp(-lpd)) # Expit(...) in Eq 28
    }
    else { #No CF adjustment
      fiter=glm(yi~Ci,family = binomial(link = "logit"))$coef
      lpd=fiter[1]+fiter[2]*lpredc
      ppred=1/(1+exp(-lpd))
    }
    predder=lm(ppred~as.factor(ldose)-1)$coef-gold #Eq 28-beta
    Out=rbind(Out, c(pred,predder))
  }
  jk=apply(Out,2,var,na.rm=T)
  jk[6:10]/jk[1:5]
}

Pout=NULL
Pout=simu(adj=0)
Pout=rbind(Pout,simu())
Pout=rbind(Pout,simu(rho=0.3))
Pout=rbind(Pout,simu(rho=0.6))
```

```r
Pout=rbind(Pout,simu(rho=0.9))

Pout=rbind(Pout,simu(ns=80,adj=0))
Pout=rbind(Pout,simu(ns=80))
Pout=rbind(Pout,simu(ns=80,rho=0.3))
Pout=rbind(Pout,simu(ns=80,rho=0.6))
Pout=rbind(Pout,simu(ns=80,rho=0.9))

dose=(1:5)
pdf(file="logit1.pdf")
par(mar=c(5.1, 4.1, 4.1, 1.1),mfrow=c(1,2))
plot(dose,ylim=c(0.4,1),Pout[1,],xlab="Dose",ylab="Var. ratio(DER to DR)",type="l",main="n=40",lty=5)
lines(dose,ylim=c(0,1),Pout[2,],type="l",lty=1)
lines(dose,ylim=c(0,1),Pout[3,],type="l",lty=2)
lines(dose,ylim=c(0,1),Pout[4,],type="l",lty=3)
lines(dose,ylim=c(0,1),Pout[5,],type="l",lty=4)
legend(1.5,0.6,legend=c(expression(paste(rho == 0, " Unadj")),
                        expression(rho == 0),expression(rho == 0.3),expression(rho == 0.6),expression(rh

plot(dose,ylim=c(0.4,1),Pout[6,],xlab="Dose",ylab="Var. ratio(DER to DR)",type="l",main="n=80",lty=5)
lines(dose,ylim=c(0,1),Pout[7,],type="l",lty=1)
lines(dose,ylim=c(0,1),Pout[8,],type="l",lty=2)
lines(dose,ylim=c(0,1),Pout[9,],type="l",lty=3)
lines(dose,ylim=c(0,1),Pout[10,],type="l",lty=4)
legend(1.5,0.6,legend=c(expression(paste(rho == 0, " Unadj")),
                        expression(rho == 0),expression(rho == 0.3),expression(rho == 0.6),expression(rh
dev.off()

# Empirical mean method based on probit models
simu=function(ns=40,rho=0,scal=1.5,nsimu=10000,adj=1){
  set.seed(123)
  nsub=200000
  dose=(1:5)/scal
  di=rep(dose, length=nsub)
  ui=rnorm(nsub)
  av=rho
  bv=sqrt(1-av^2)
  Ci=di+av*ui+bv*rnorm(nsub)-3
  yi=ifelse(Ci+av*ui+bv*rnorm(nsub)>0,1,0)
  gold=lm(yi~as.factor(di)-1)$coef

  nsub=ns
  Out=NULL
  for (j in 1:nsimu){
    di=rep(dose, length=nsub)
    ui=rnorm(nsub)
    Ci=di+av*ui+bv*rnorm(nsub)-3
    yi=ifelse(Ci+av*ui+bv*rnorm(nsub)>0,1,0)
    fitdr=glm(yi~di,family = binomial(link = "probit"))
    pred=pnorm(fitdr$coef[1]+fitdr$coef[2]*dose)-gold
    fitde=lm(Ci~di)
```

```r
      predc=fitde$coef[1]+fitde$coef[2]*dose
      ldose=rep(dose,rep(nsub,length(dose)))
      vi=Ci-predc
      lvi=rep(vi,length(dose))
      lpredc=fitde$coef[1]+fitde$coef[2]*ldose+lvi
      if (adj==1){
         fiter=glm(yi~Ci+vi,family = binomial(link = "probit"))$coef
         lpd=fiter[1]+fiter[2]*lpredc+fiter[3]*lvi
         ppred=pnorm(lpd)
      }
      else {
         fiter=glm(yi~Ci,family = binomial(link = "probit"))$coef
         lpd=fiter[1]+fiter[2]*lpredc
         ppred=pnorm(lpd)
      }
      predder=lm(ppred~as.factor(ldose)-1)$coef-gold
      Out=rbind(Out, c(pred,predder))
   }
   jk=apply(Out,2,var,na.rm=T)
   c(jk[6:10]/jk[1:5])
}

Pout=NULL
Pout=simu(adj=0)
Pout=rbind(Pout,simu())
Pout=rbind(Pout,simu(rho=0.3))
Pout=rbind(Pout,simu(rho=0.6))
Pout=rbind(Pout,simu(rho=0.9))

Pout=rbind(Pout,simu(ns=80,adj=0))
Pout=rbind(Pout,simu(ns=80))
Pout=rbind(Pout,simu(ns=80,rho=0.3))
Pout=rbind(Pout,simu(ns=80,rho=0.6))
Pout=rbind(Pout,simu(ns=80,rho=0.9))

dose=(1:5)/1.5
pdf(file="probit15.pdf")
par(mar=c(5.1, 4.1, 4.1, 1.1),mfrow=c(1,2))
plot(dose,ylim=c(0.4,1),Pout[1,],xlab="Dose",ylab="Var. ratio(DER to DR)",type="l",main="n=40",lty=5)
lines(dose,ylim=c(0,1),Pout[2,],type="l",lty=1)
lines(dose,ylim=c(0,1),Pout[3,],type="l",lty=2)
lines(dose,ylim=c(0,1),Pout[4,],type="l",lty=3)
lines(dose,ylim=c(0,1),Pout[5,],type="l",lty=4)
legend(1.5,0.6,legend=c(expression(paste(rho == 0, " Unadj")),
                        expression(rho == 0),expression(rho == 0.3),expression(rho == 0.6),expression(

plot(dose,ylim=c(0.4,1),Pout[6,],xlab="Dose",ylab="Var. ratio(DER to DR)",type="l",main="n=80",lty=5)
lines(dose,ylim=c(0,1),Pout[7,],type="l",lty=1)
lines(dose,ylim=c(0,1),Pout[8,],type="l",lty=2)
lines(dose,ylim=c(0,1),Pout[9,],type="l",lty=3)
lines(dose,ylim=c(0,1),Pout[10,],type="l",lty=4)
```

```
legend(1.5,0.6,legend=c(expression(paste(rho == 0, " Unadj")),
                        expression(rho == 0),expression(rho == 0.3),expression(rho == 0.6),expression(
dev.off()
```